%% file: proceedings_ef.tex
\documentclass[graybox]{svmult}

\usepackage{mathptmx}       
\usepackage{helvet}         
\usepackage{courier}        
\usepackage{type1cm}        
\usepackage{makeidx}         
\usepackage{graphicx}        
\usepackage{multicol}        
\usepackage[bottom]{footmisc}

\makeindex             

\begin{document}

\title*{Precise Measurement of \\ the {\boldmath$\pi^+\to {e}^+\nu$} Branching Ratio}

\author{
E.~Frle\v{z},$^{\rm a}$
L.~P.~Alonzi,$^{\rm a}$
V.~A.~Baranov,$^{\rm b}$
W.~Bertl,$^{\rm c}$
M.~Bychkov,$^{\rm a}$
Yu.M.~Bystritsky,$^{\rm b}$
N.V.~Khomutov,$^{\rm b}$
A.S.~Korenchenko,$^{\rm b}$
S.M.~Korenchenko,$^{\rm b}$
M.~Korolija,$^{\rm d}$
T.~Kozlowski,$^{\rm e}$
N.P.~Kravchuk,$^{\rm b}$
N.A.~Kuchinsky,$^{\rm b}$
D.~Mekterovi\'c,$^{\rm d}$
D.~Mzhavia,$^{\rm {b,f}}$
A.~Palladino,$^{\rm a}$
D.~Po\v{c}ani\'c,$^{\rm a}$
P.~Robmann,$^{\rm g}$
A.M.~Rozhdestvensky,$^{\rm b}$
V.V.~Sidorkin,$^{\rm b}$
U.~Straumann,$^{\rm g}$
I.~Supek,$^{\rm d}$
P. Tru\"ol,$^{\rm g}$
A.~van~der~Schaaf,$^{\rm g}$
E.P.~Velicheva,$^{\rm b}$ and
V.P.~Volnykh$^{\rm b}$
}
\institute{Emil Frle\v{z}, ph.: 434-924-6786, \hbox{\email{frlez@virginia.edu}}.
{\at
 $^{\rm a}$Department of Physics, University of Virginia, Charlottesville, VA 22904-4714,
 USA}
{\at
 $^{\rm b}$Joint Institute for Nuclear Research, RU-141980 Dubna, Russia}
{\at
 $^{\rm c}$Paul Scherrer Institut, CH-5232 Villigen PSI, Switzerland}
{\at
 $^{\rm d}$Rudjer Bo\v{s}kovi\'c Institute, HR-10000 Zagreb, Croatia}
{\at
 $^{\rm e}$Institute for Nuclear Studies, PL-05-400 Swierk, Poland}
{\at
 $^{\rm f}$IHEP, Tbilisi, State University, GUS-380086 Tbilisi, Georgia}
{\at
 $^{\rm g}$Physik Institut der Universit\"at Z\"urich, CH-8057 Z\"urich,
 Switzerland}
     }
%
%
\maketitle

\abstract{The PEN Collaboration is conducting a new measurement of the
$\pi^+\to e^+\nu$ branching ratio at the Paul Scherrer Institute, with the goal uncertainty of
$\delta B/B_{\pi e2}=5\times 10^{-4}$ or lower. At present, the combined accuracy of
all published $\pi_{e2}$ decay measurements lags behind the theoretical calculation by 
a factor of 40. In this contribution we report on the PEN detector configuration and its performance 
during two development runs done in 2007 and 2008.}


\section{Introduction}\label{sec:intr}

The PEN Collaboration is performing a new measurement of the $\pi^+\to e^+\nu (\gamma)$
branching ratio ($B_{\pi e2}$) with a relative uncertainty of $\sim 5\times 10^{-4}$ or lower, at 
the Paul Scherrer Institute (PSI)~\cite{pen}. 

The amplitude for this rare pion decay is a textbook example of the $V-A$ nature of 
the electroweak interaction with the branching ratio understood at the level of 
less than one part in $10^4$.

Recent independent theoretical calculations of the ratio of the decay rates 
$\Gamma(\pi\to e\bar{\nu}(\gamma))/\Gamma(\pi\to \mu\bar{\nu}(\gamma))$
are in a very good agreement and give,
\begin{eqnarray}
B^{\rm SM}_{e/\mu}=\frac{\Gamma(\pi\to e\bar{\nu}(\gamma))}{\Gamma(\pi\to \mu\bar{\nu}(\gamma))}{|}_{\rm the}=
\cases{ (1.2352\pm 0.0005)\times 10^{-4},& Ref.~\cite{Mar93},\cr
        (1.2354\pm 0.0002)\times 10^{-4},& Ref.~\cite{Fin96},\cr
        (1.2352\pm 0.0001)\times 10^{-4},& Ref.~\cite{Cir07}. }\cr
\label{eq:1}
\end{eqnarray}
Work of Marciano and Sirlin's~\cite{Mar93} and Finkemeier~\cite{Fin96} took into account radiative corrections, 
higher order electroweak leading logarithms, short-distance QCD corrections and structure-dependent effects, 
while Cirigliano and Rosell~\cite{Cir07} arrived at the value using the two-loop Chiral Perturbation Theory.

The two most recent competitive experiments completed 15 years ago
at the TRIUMF cyclotron~\cite{Bry92} and the PSI ring accelerator~\cite{Cza93} are also mutually
consistent, but are exceeded in accuracy by a factor of 40 by the latest theoretical calculations~(Eq.~\ref{eq:1}):
\begin{eqnarray}
B^{\rm exp}_{e/\mu}=
\cases{ (1.2265\pm 0.0034({\rm stat})\pm 0.0044({\rm sys}) )\times 10^{-4},& Ref.~\cite{Bry92},\cr
        (1.2350 \pm 0.0035({\rm stat})\pm 0.0036({\rm sys}) )\times 10^{-4},& Ref.~\cite{Cza93}. }\cr
\end{eqnarray}
These two $\pi_{e2}$ measurements, though inadequate in precision when compared with the theoretical accuracy, 
represent the best experimental test of $\mu$-$e$ universality at present. The improved measurement 
would, in addition, be a very sensitive probe 
of all Standard Model extensions that induce pseudoscalar currents~\cite{Bry93} and serve as a test of
possible supersymmetric corrections to the lepton couplings~\cite{Mas06}.

\section{Experimental Apparatus}\label{sec:appar}

\begin{figure}[b]
\hbox{
\includegraphics[scale=.60]{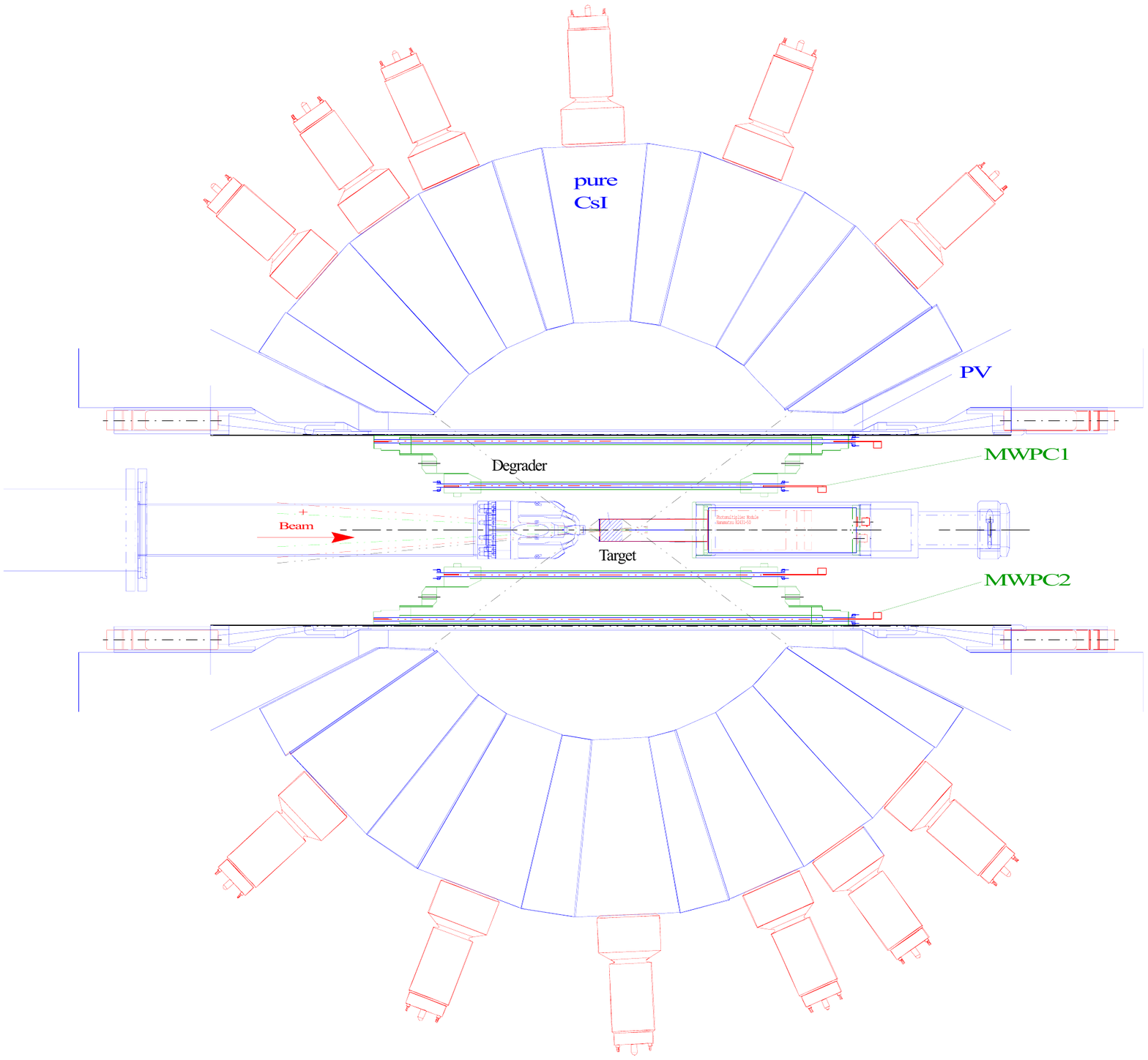}}
\vglue -7.0cm
\hbox{\hglue -0.5cm
\includegraphics[scale=.60]{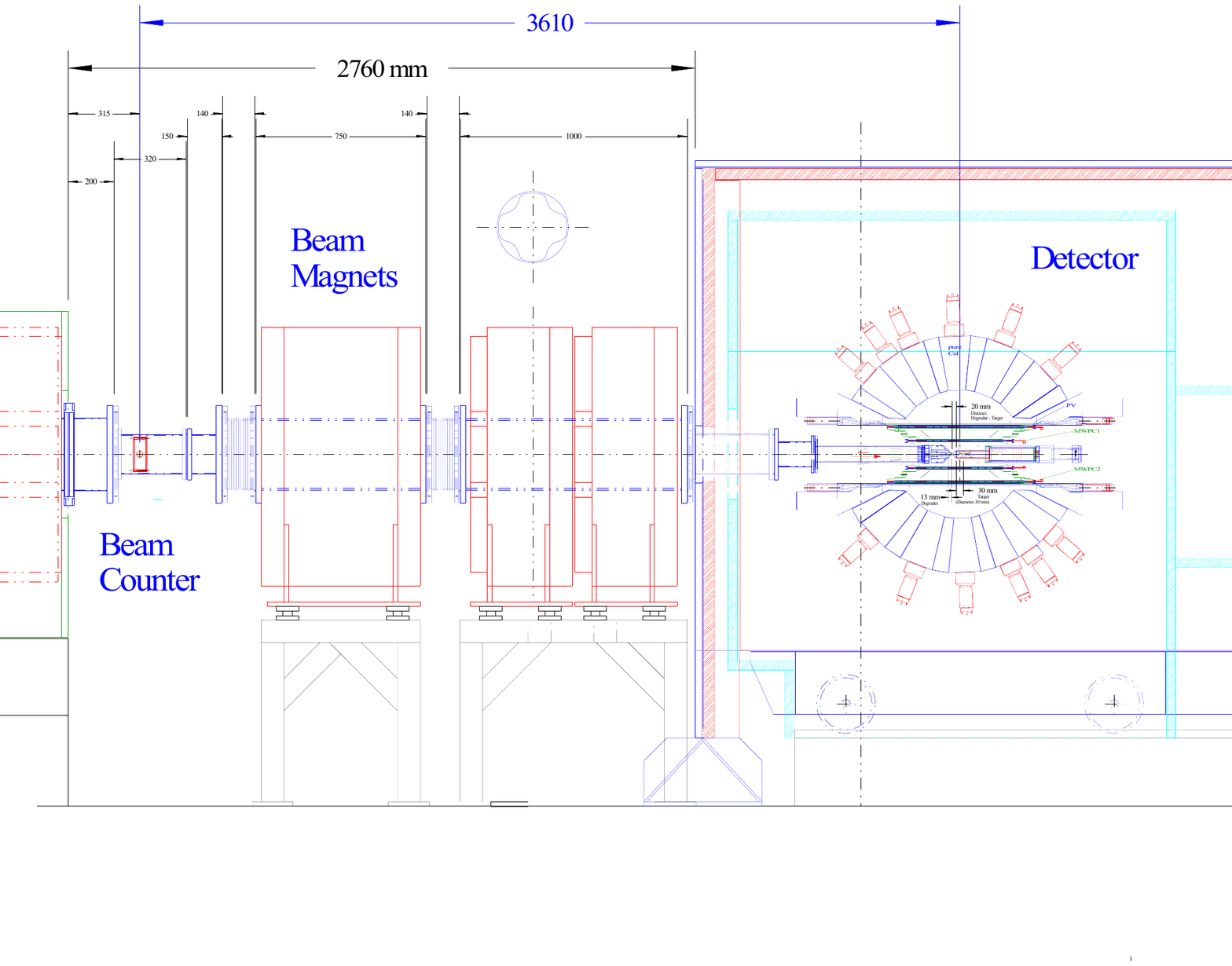}}
\vglue -7.5cm
\caption{Cross sections through the PEN set up in the 2008
configuration. Top panel shows the central tracking region,
including the 4-wedge tracking degrader, active target,
segmented plastic scintillator, a pair of cylindrical MWPC's,
and modular CsI calorimeter.
Bottom panel depicts the detector in the
experimental area with the upstream beam-tagging counter inside the vacuum
pipe and the triplet of focusing quadrupole magnets.}
\label{fig:det}       
\end{figure}

The PIBETA detector, used from 1999 to 2004 in a series of rare pion and muon decay measurements~\cite{pb1,pb2,pb3,byc08},
was upgraded to meet the needs of the new experiment.
The {\sc PEN} apparatus is a large solid angle non-magnetic detector optimized
for detection of photons and electrons emanating from the pion and muon decays in the
centrally located stopping target.
The main detector elements of the apparatus, shown in two panels of Fig.~\ref{fig:det}, are:
\begin{itemize}
\item[(1)]\ a thin upstream beam counter,
and an active degrader all made of plastic scintillator material, used for the beam definition;
\item[(2)]\ an active plastic scintillator target, used to stop and detect the beam particles,
and record their charged decay products;
\item[(3)]\ two concentric low-mass cylindrical multi-wire proportional chambers 
for charged particle tracking, surrounding the active target;
\item[(4)]\ a 20-piece fast plastic scintillator hodoscope, surrounding the MWPC's,
used for particle identification and fast timing;
\item[(5)]\ a high-resolution segmented fast shower CsI calorimeter, surrounding
the target region and tracking detectors in a near-spherical geometry;
\end{itemize}

The electromagnetic calorimeter is made of 240 pure CsI scintillator modules covering a $\sim 3\pi$ 
sr solid angle, with openings
allowing for beam entry and detector readout. The inner radius of the hollowed CsI sphere is
26\,cm and its thickness of 22\,cm corresponds to 12 radiation lengths.

The key element of the fast electronic logic is the one-arm calorimeter trigger defined by
discriminated analog signal sum of any one or more of 60 groups of 9 CsI calorimeter modules. 
The high discrimination threshold (HT) is adjustable and was set at around 44\,MeV in 
the PEN experiment. In parallel, we have also used the pre-scaled (1:16 or 1:64) low level
threshold (LT) trigger set by a discriminated sum of the plastic
scintillator hodoscope signals. The calorimeter energy spectrum for LT trigger 
extends well below 0.5\,MeV, being limited only by the ADC pedestal widths of the individual
CsI detectors.

The PEN experiment completed two development runs in 2007 and 2008.
The total number of recorded $\pi^+$ stops was $8.1\times 10^{10}$, while the total number
of collected triggers was $4.7\times 10^6$ for HT and 180,000 for LT events.

\section{Experimental Method}\label{sub:meth}

The $\pi_{e2}$ events are collected primarily by means of the one-arm high-threshold 
calorimeter trigger. The threshold value was chosen so as to minimize 
the fraction of events in the $\pi^+$ energy spectrum ``tail''
caused by the electromagnetic shower leakage in the CsI calorimeter,
while keeping the data acquisition live time fraction at $\sim 0.8-0.9$.
 
The $\pi\to e\nu$ branching ratio can be evaluated as
\begin{equation}
B_{\pi e2}=\frac{N_{\rm HT}(1+\epsilon)}{A_{\pi e2} N_{\pi^+}f_{\pi e2}(T)},
\end{equation}
where $N_{\rm HT}$ is the number of recorded $\pi_{e2}$ events above the $E_{\rm HT}$
energy, $\epsilon=N_t/N_{\rm HT}$ is the ratio of the ``tail'' to ``HT''  $\pi_{e2}$ events,
$A_{\pi e2}$ is the detector acceptance, $N_{\pi^+}$ is the number of stopped beam pions 
recorded during the experiment, while $f_{\pi e2}(T)$ is the pion decay probability between the pion stop
time, $t=0$, and the end of the trigger gate, $t=T\simeq 220\,$ns. The most precise stopped pion count
is obtained by recording the sequential (Michel) $\pi\to\mu\to e$ chain decay:
\begin{equation}
B_{\pi\mu e}=\frac{N_{\pi\mu e}}{A_{\pi\mu e}N_{\pi^+}f_{\pi\mu e}(T) },
\end{equation}
where $B_{\pi\mu e}\simeq 1$ is the branching ratio for the $\mu\to e\nu\bar{\nu}(\gamma)$ decay,
$A_{\pi\mu e}$ and $N_{\pi\mu e}$ are the detector acceptance and the number of muon decays, and
$ f_{\pi\mu e}$ is the trigger gate decay probability.

Combining the two expressions to eliminate $N_{\pi^+}$ gives:
\begin{equation}
B_{\pi e2}=\frac{N_{\rm HT}(1+\epsilon)}{N_{\pi\mu e}}\cdot \frac{A_{\pi\mu e}}{A_{\pi e2}}
\cdot\frac{f_{\pi\mu e}(T)}{f_{\pi e2}(T)}B_{\pi\mu e},
\end{equation}
which conveniently factorizes into quantities that share many of the same systematic uncertainties.

A detailed analysis of the optimum choice of trigger parameters ($T$, $E_{\rm HT}$, 
LT/HT prescaling, DAQ rates, etc.), presented in~Ref.\cite{pen}, has demonstrated
that the PEN detector system is capable of reaching statistical uncertainty levels in $\pi_{e2}$
decay that are an order of magnitude better than those obtained in previous experiments
in about six months of beam time.

The systematic uncertainties relevant to the PEN measurements are:
\begin{itemize}
\item[(1)]\ discrimination of pion and muon events: due to low $\mu$ decay pile-up and the 
digitized target waveforms misidentified events will be kept at the level below $10^{-4}$.
\item[(2)]\ the pion and muon decay normalization: the well determined Michel parameter $\rho$
controlling the shape of the $\mu^+$ decay positron energy spectrum, in conjunction
with the low energy threshold below 1\,MeV, and the absolute calorimeter energy calibration 
attained in previous measurements, are projected to yield $\Delta N_{\pi\mu e}/N_{\pi\mu e}=1\times 10^{-4}$. 
\item[(3)]\ the ratio of acceptances for $\pi_{e2}$ and Michel decay events: shared systematics
of the signal and normalization decay limits the uncertainties to $\sim 1\times 10^{-4}$.
\item[(4)]\ a correction for radiative muon decays: in-situ measurement of the radiative
decays leads to sub-$10^{-4}$ accuracy.
\item[(5)]\ nuclear interactions correction: suppressed in the HT data sample via analysis cuts, 
will be simulated for the LT ``tail'' in the full detector Monte Carlo calculation to 10\,\% accuracy.
It will also be measured with our LT trigger. 
\item[(6)]\ the zero time definition: 5\,ps accuracy in the mean value for $t=0$ 
is achievable with digitized waveforms.
\end{itemize}
Thus, an overall systematic uncertainty in the range of $2-4\times 10^{-4}$ is an attainable 
goal in the PEN experiment.

\section{Preliminary Results}\label{sec:cent}

Tightly focused $\pi^+$ beam tunes were developed for a dozen beam momenta between
67 and 85\,MeV/c, with the pion stopping rates ranging from 1,000 to 20,000\,$\pi^+$/s.
The beam particles are first detected in a thin upstream beam counter made of plastic scintillator, 
3.61\,m upstream of the detector center. The beam pions are 
subsequently slowed down in the active degrader and stopped in the active target.

We have used several cylindrical-shaped degraders of different 
thicknesses, the final one being a 4-wedge scintillator capable of lateral
beam particle tracking. It consisted of two pairs of plastic scintillators with the
thicknesses tapering from 5 to 1.5\,mm along the horizontal and vertical directions.
The lateral beam particle coordinates were determined by the ratios of energy depositions
in the pairs of wedges. The points of closest approach (PCA) of the reconstructed beam particle
paths and the back-tracked decay positron hits recorded by the MWPC's define
the $\pi^+$ or $\mu^+$ decay vertex inside the target. Comparison of
a Monte Carlo simulation of the PCA distributions with the experimental  
histograms reveals that the vertex resolution root-mean-square is better than 2\,mm.

The reconstructed $\pi^+/\mu^+$ vertex allows, in turn, the calculation of the $e^+$
path length in the target and the prediction of the positron contribution to the
measured target signal.

The PMT signals for the upstream beam detector, active degrader and active
target were digitized in conventional FASTBUS and CAMAC ADC and TDC units as well
as by a waveform digitizer, the Acqiris DC282 four-channel unit operated
at 2\,GS/s sampling rate (2 channels/ns), and yielding an effective 7-bit resolution. 

\begin{figure}[b]
\sidecaption
\bigskip
\hbox to \textwidth{
\includegraphics[width=0.54\textwidth]{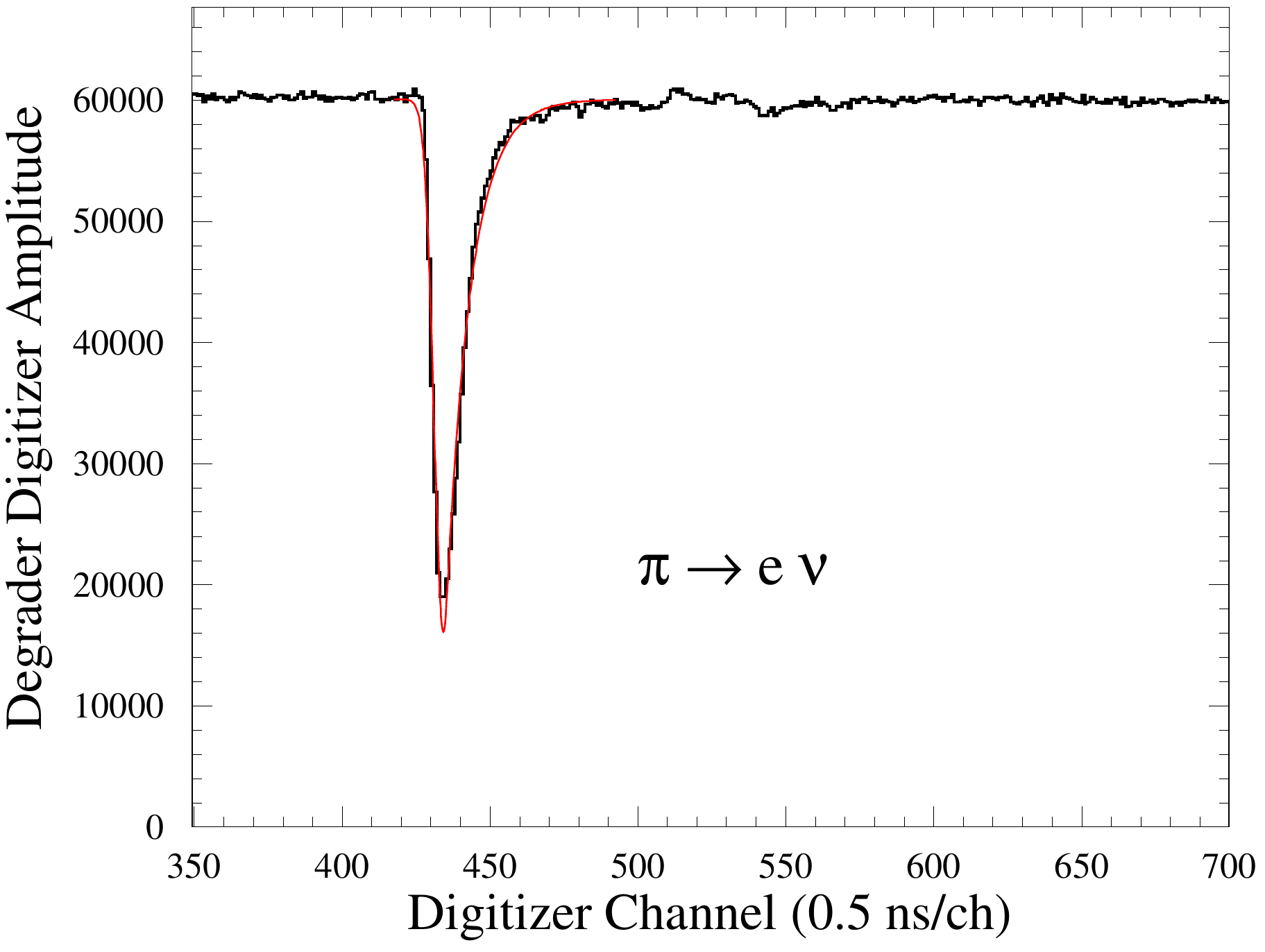}
\includegraphics[width=0.54\textwidth]{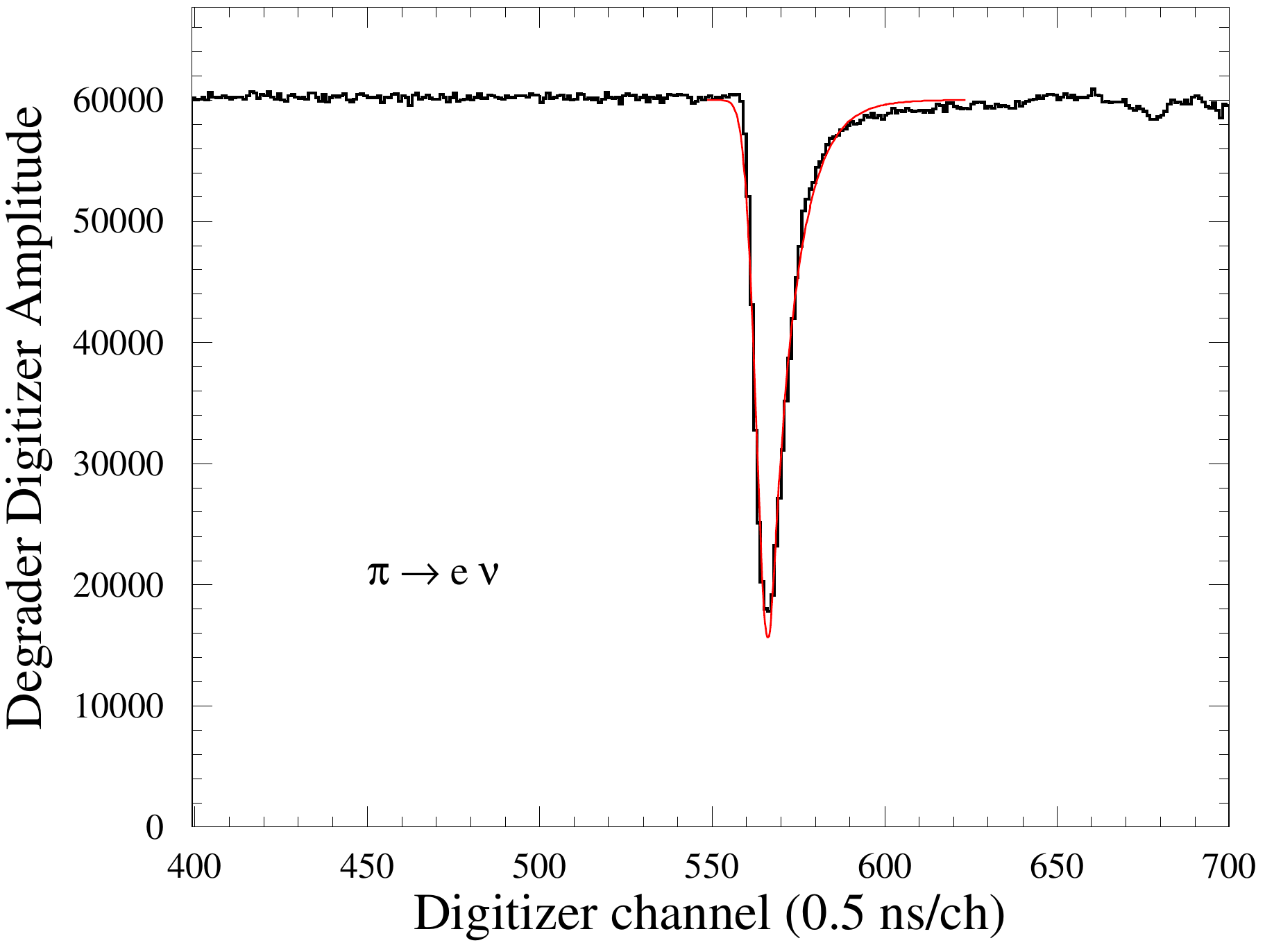}
}
\hbox to \textwidth{
\includegraphics[width=0.54\textwidth]{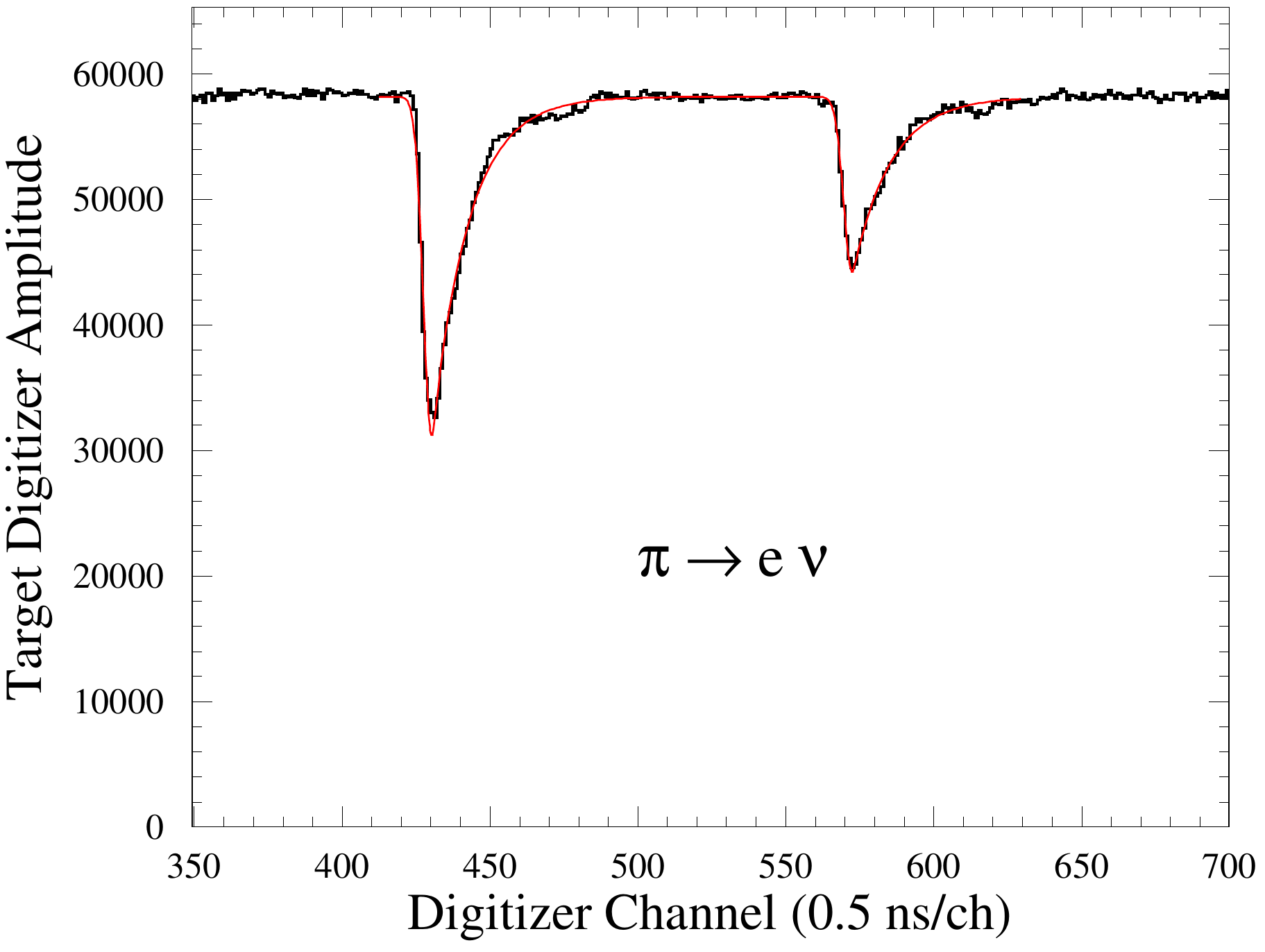}
\includegraphics[width=0.56\textwidth]{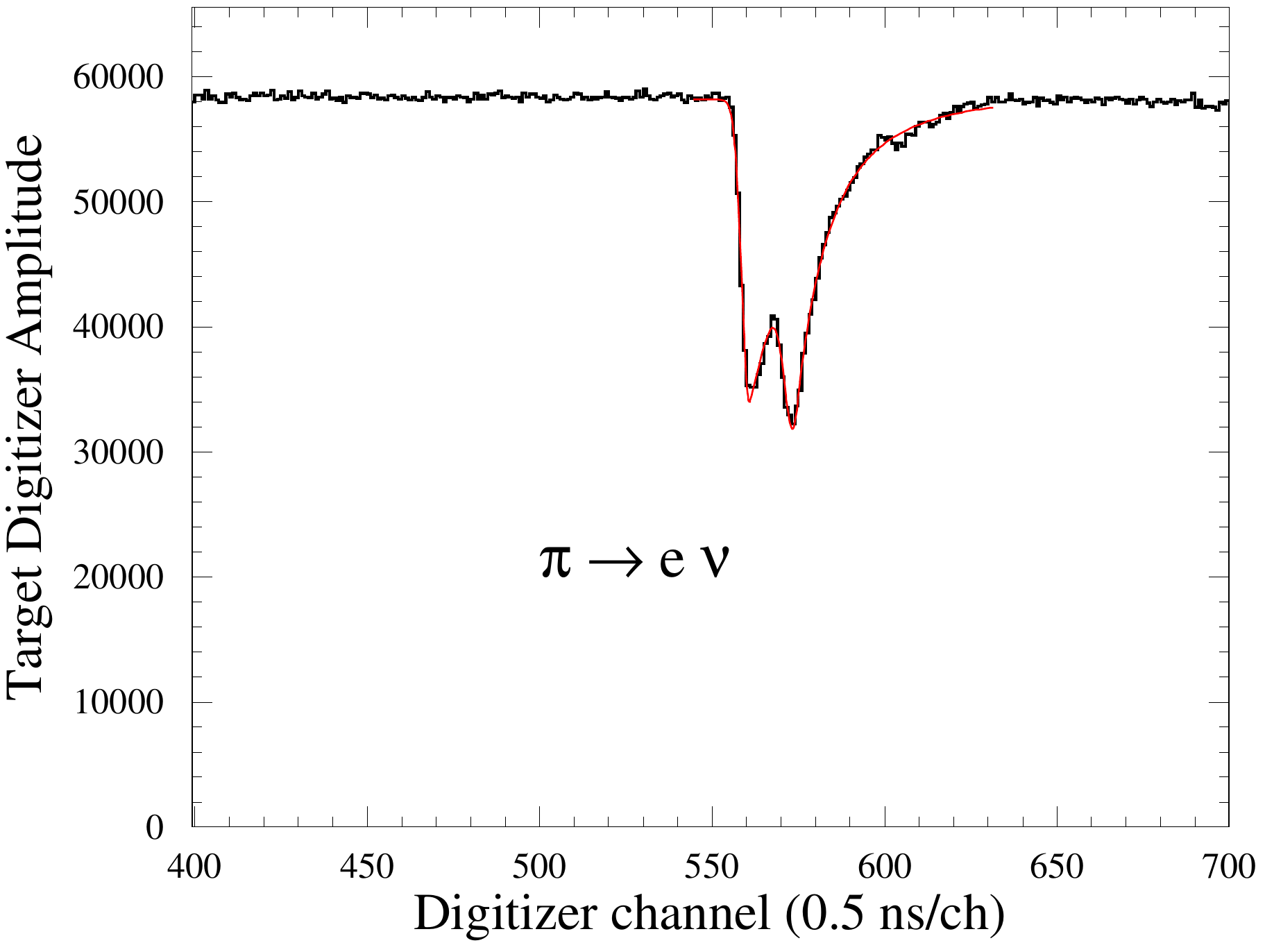}
}
\hbox to \textwidth{
\includegraphics[width=0.54\textwidth]{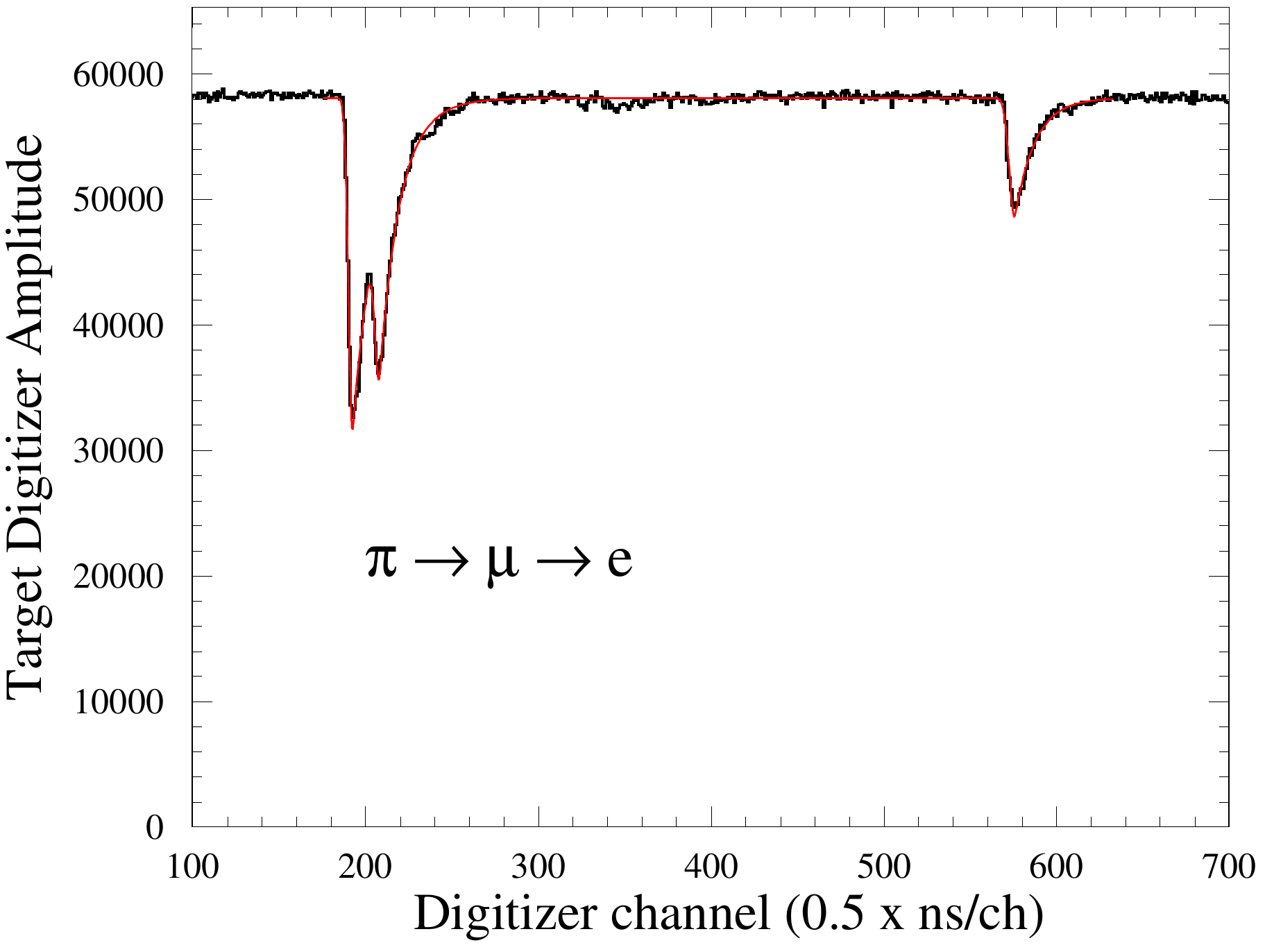}
\includegraphics[width=0.54\textwidth]{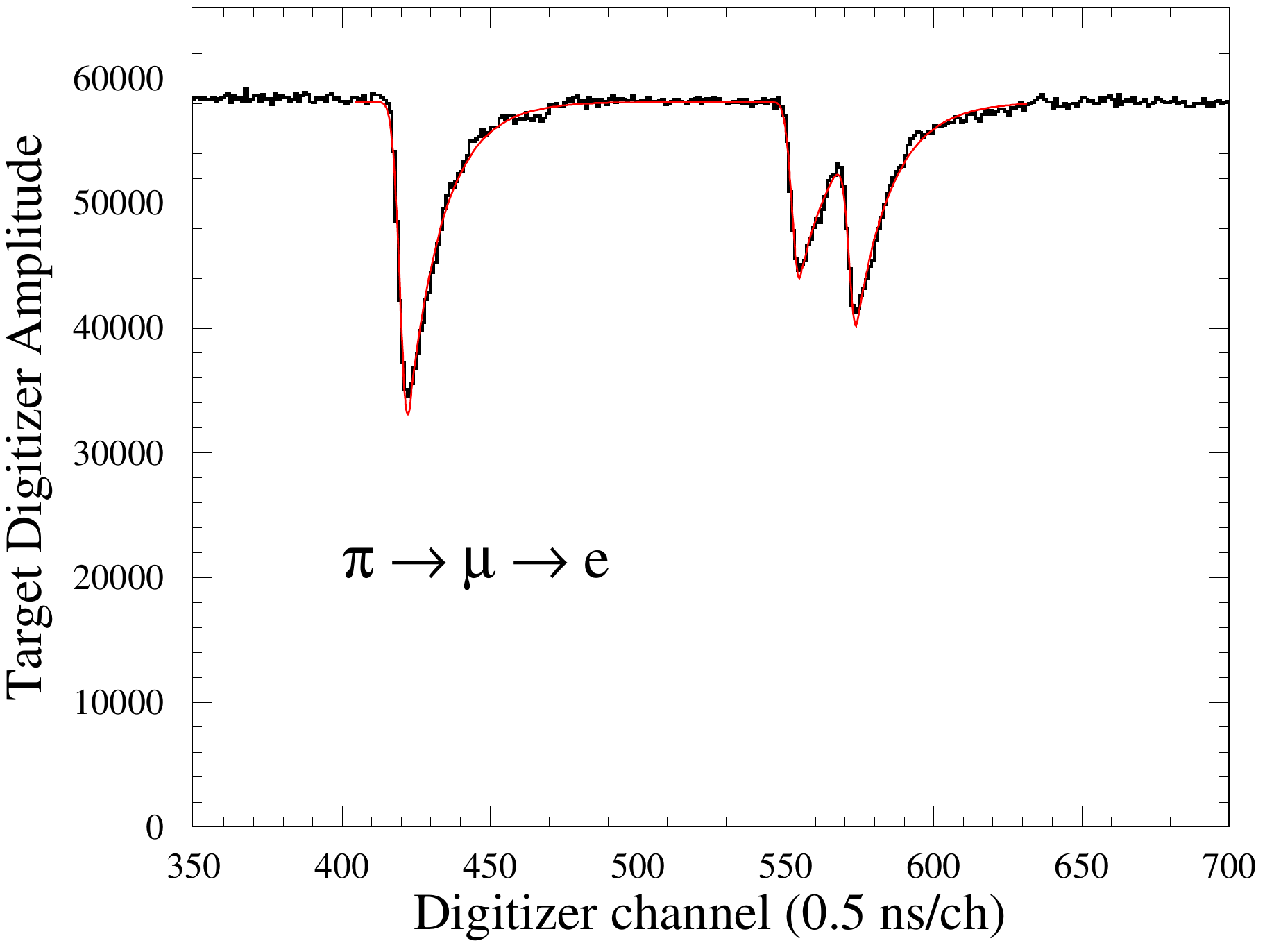}
}
\caption{Digitized PMT waveforms of the single-piece active degrader
counter (top panels) and the corresponding active target waveforms for
two $\pi\to e\nu$ decay events (middle panels) and two Michel 
$\pi\to\mu\to e$ chain decay events (bottom panels) from the 2007 development run. 
The time scale is 2 channels/ns.}
\label{fig:wave}
\end{figure}

\begin{figure}[b]
\vglue -8.5cm
\hbox{
\includegraphics[scale=.66]{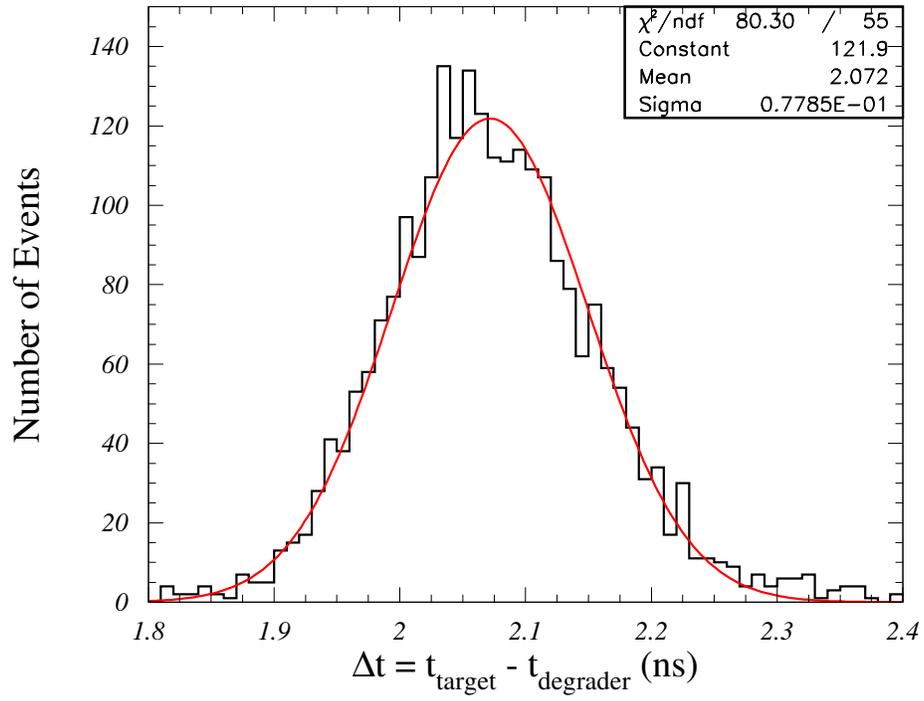} }
\caption{
Time-of-flight spectrum between the degrader and target with {\sl rms}=78\,ps. 
The contributions to {\sl rms} come from the beam momentum spread and instrumental resolution.}
\label{fig:timeres}       
\end{figure}

Examples of the recorded beam counter waveforms are given in the six
panels of Fig.~\ref{fig:wave}. The data quality is illustrated in the $\pi^+$ time-of-flight
(TOF) spectrum (Fig.~\ref{fig:timeres}), representing the time differences between the
target and the degrader hits.

\begin{figure}[b]
\vglue -8.75cm
\hbox{
\includegraphics[scale=.64]{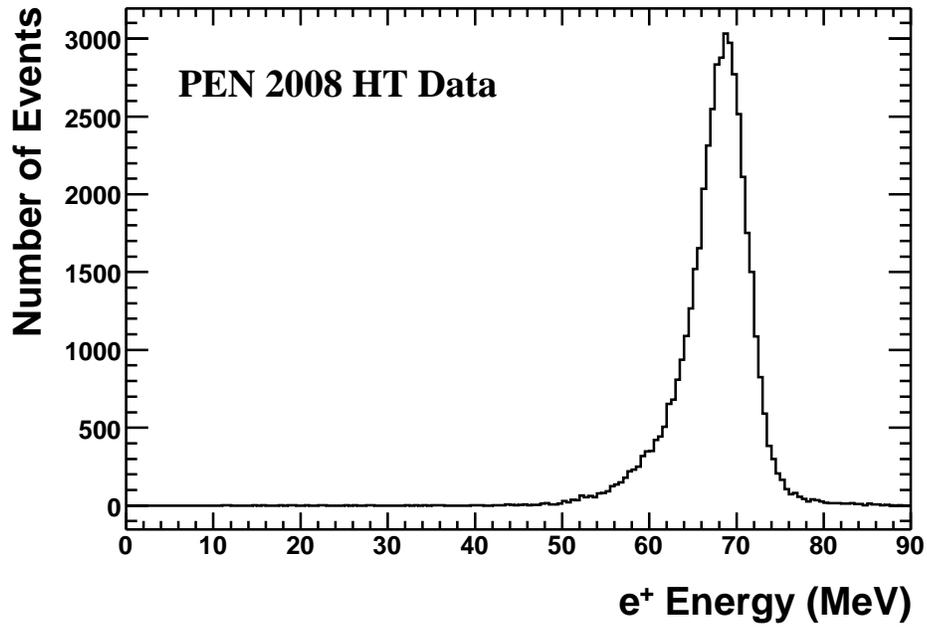} }
\caption{The high threshold $\pi^+\to e^+\nu$ energy spectrum for a subset of the 2008 data.
Michel background has been suppressed by the cut on the total energy
deposited in the target.}
\label{fig:lineshape}       
\end{figure}

By applying a cut on the monoenergetic 4.1\,MeV $\mu^+$ peak in the
active target waveform, one can discriminate between the $\pi_{e2}$ signal events
and $\pi\to\mu\to e$ background events after the $e^+$-in-target energy contribution
has been subtracted. The total energy spectrum of the $\pi_{e2}$ decay 
positron identified by such a cut is shown in Fig.~\ref{fig:lineshape} for
a subset of 2008 HT data.    

\begin{figure}[b]
\sidecaption
\bigskip
\includegraphics[scale=.64]{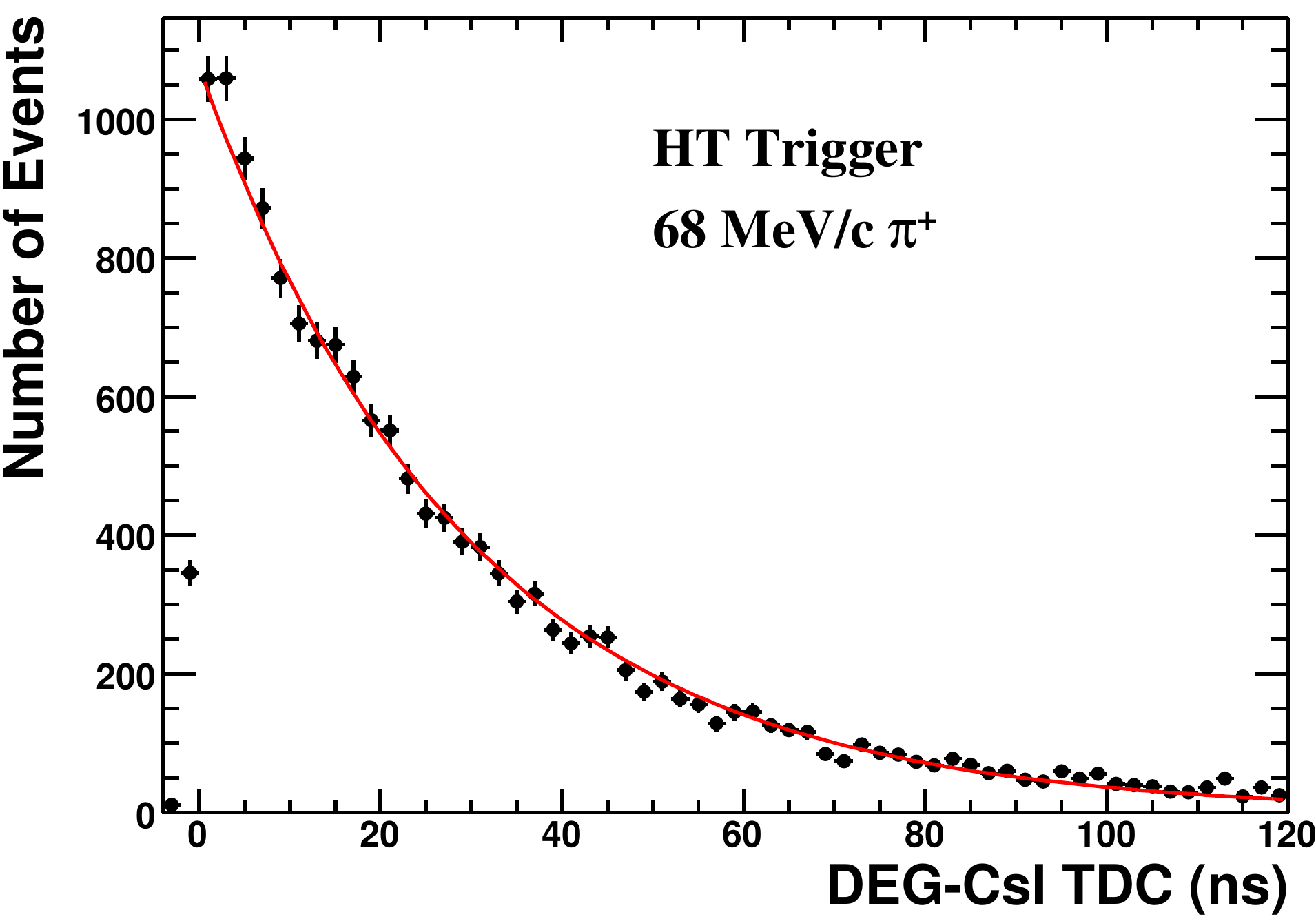}
\includegraphics[scale=.64]{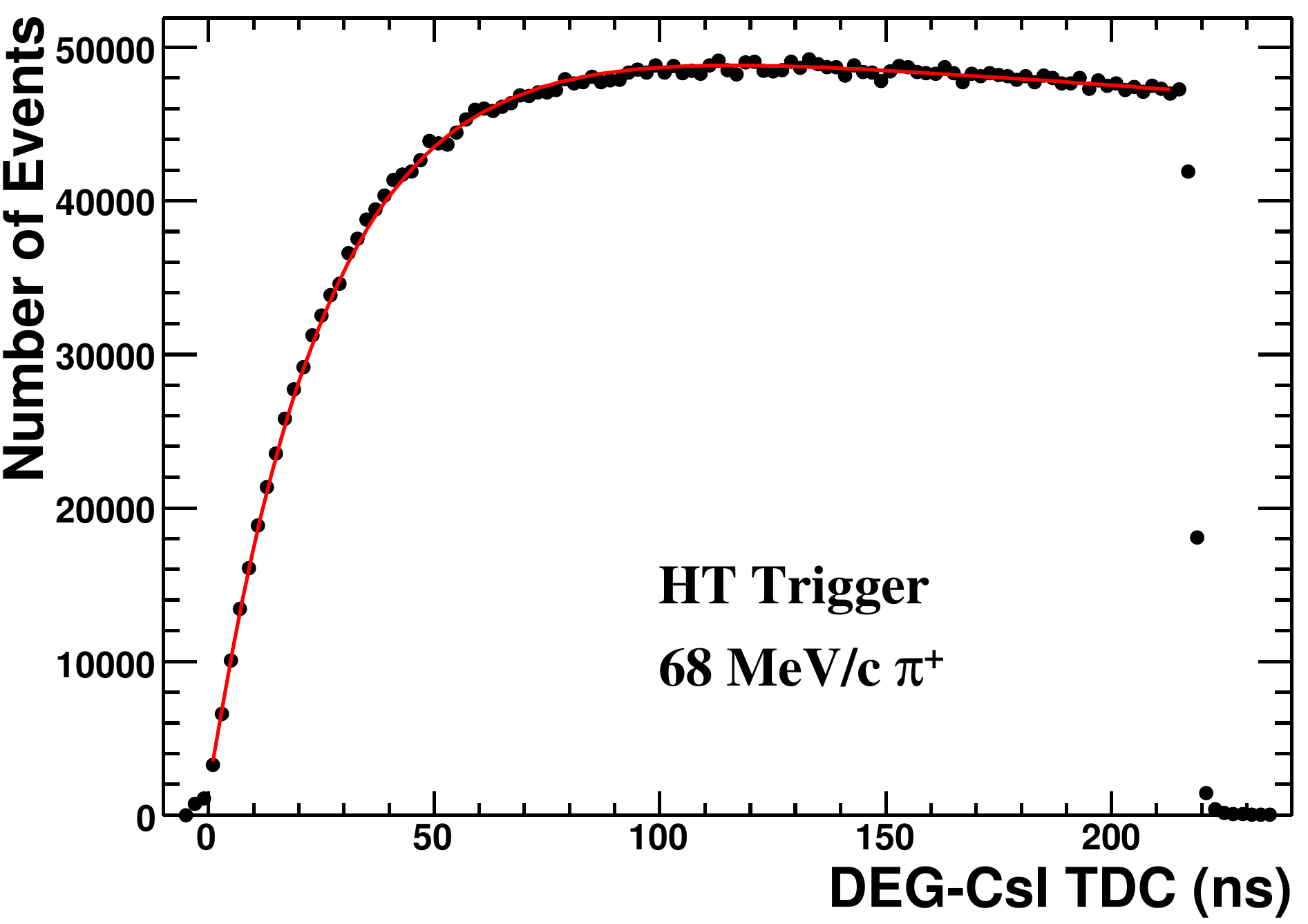}
\caption{Calorimeter time distributions with respect to the 
$\pi^+$ stop time for $\pi\to e\nu$ events (top panel) and
$\pi\to\mu\to e$ decay chain events (bottom panel) for 
a subset of 2007 data.}
\label{fig:time}       
\end{figure}

Finally, the time spectra of the $\pi_{e2}$ signal and Michel background events with
respect to the stopping $\pi^+$ time are shown in the Fig.~\ref{fig:time}. The $rms$
time resolution between the target and degrader is 78\,ps for beam pions. Similarly,
the $rms$ time resolution between the upstream beam detector and the active degrader is 
approximately 120\,ps. These time resolutions correspond to $O(10\,\%)$ of the bin
width of the waveforms from which the timing was deduced.

\section{Conclusion and Future Plans}\label{sec:conc}

We have upgraded the PIBETA detector to optimize it for the task of
a precise measurement of the $\pi^+\to e^+\nu$ decay ratio at PSI .
Two development runs were successfully completed in 2007 and 2008,
with the beam stop and DAQ rates ramped up to the design specifications.
To date we have recorded $4.7\times 10^6$ raw $\pi\to e\nu$ events, before
analysis cuts are applied, corresponding
to the statistical uncertainty of $\delta B/B=5\times 10^{-4}$.
The replay of the data set collected so far is under way as of this writing, 
in preparation for a production run in 2009 that is planned to complete 
the required statistics~\cite{www}.

The PEN experiment has been supported by the National Science Foundation
(NSF PHY-0653356), the Paul Scherrer Institute and the Russian Foundation for 
Basic Research (Grant No. 08-02-00652a).

\input{references}

\end{document}

%% file: references.tex
%
%
%